\documentclass[10pt,preprint2]{aastex}

\def \apj {ApJ}
\def \apjs {ApJS}
\def \apjl {ApJ}
\def \solphys {Solar Phys.}

\def \aap {A\&A}

\newcommand{\citeN}[1]{\citeauthor{#1} (\citeyear{#1})}
\newcommand{\citeNP}[1]{\citeauthor{#1} \citeyear{#1}}

\newcommand{\FeI}{\ion{Fe}{1}}
\newcommand{\OI}{\ion{O}{1}}

\newcommand{\lgEO}{$\log \epsilon_{\rm O}$}

\shortauthors{Socas-Navarro \& Norton}
\shorttitle{The Solar Oxygen Crisis}

\begin{document}

\title{The solar oxygen crisis: Probably not the last word}

\author{H. Socas-Navarro}
   	\affil{High Altitude Observatory, NCAR\thanks{The National Center
	for Atmospheric Research (NCAR) is sponsored by the National Science
	Foundation.}, 3080 Center Green Dr, Boulder, CO 80301, USA}
	\email{navarro@ucar.edu}

\author{A. A. Norton} 
   	\affil{National Solar Observatory, 950 N. Cherry Ave., Tucson,
   	AZ 85719, USA}
	\email{norton@nso.edu}

\date{}%

   
               
\begin{abstract}
In this work we present support for recent claims that advocate a
downward revision of the solar oxygen abundance. Our analysis employs
spatially-resolved spectro-polarimetric observations including the
\FeI \ lines at 6302~\AA \, and the \OI \ infrared triplet around
7774~\AA \, in the quiet Sun. We used the \FeI \ lines to reconstruct
the three-dimensional thermal and magnetic structure of the
atmosphere. The simultaneous \OI \ observations were then employed to
determine the abundance of oxygen at each pixel, using both LTE and
non-LTE (NLTE) approaches to the radiative transfer.  In this manner,
we obtain values of \lgEO=8.63 (NLTE) and 8.93 (LTE) dex.  We find an
unsettling fluctuation of the oxygen abundance over the field of
view. This is likely an artifact indicating that, even with this
relatively refined strategy, important physical ingredients are still
missing in the picture. By examining the spatial distribution of the
abundance, we estimate realistic confidence limits of approximately
0.1 dex.
\end{abstract}


Oxygen is the third most abundant chemical element in the universe.
Its abundance in the solar atmosphere has been traditionally accepted
to be approximately \lgEO=8.93 (\citeNP{AG89}), with only minor
corrections until recently (e.g., \lgEO=8.83 by \citeNP{GS98}). These
determinations are based on the use of one-dimensional (1D)
semiempirical models (such as those of \citeNP{HM74}; \citeNP{GNK+71};
\citeNP{VAL81}) to fit spectral features produced by oxygen atoms and
molecules. The most common abundance indicators are: the forbidden
[\OI] line at 6300~\AA \ (e.g., \citeNP{L78}; \citeNP{APLA01}), the
infrared (IR) 
triplet around 7774~\AA \ (e.g., \citeNP{AGS+04}; \citeNP{STBA05}), as
well as molecular bands of OH (e.g., \citeNP{AGS+04}; \citeNP{M04})
and CO (e.g., 
\citeNP{APK06}; \citeNP{SAG+06}).

Controversy was sparked when \citeN{AGS+04} applied a new hydrodynamic
simulation of the solar granulation to the determination of the oxygen
abundance, resulting in a considerably lower value
\footnote{ Although
a low abundance had been previously derived on \citeN{APLA01} based on
the analysis of the forbidden line at 6300 \AA .  } 
of \lgEO=8.66. The
new value presents some advantages, such as better agreement between
various indicators (IR triplet, [\OI] and OH vibrational and
roto-vibrational bands). It also leads to a better fit of the solar
chemical composition within its galactic environment (see, e.g.,
\citeNP{SM01} and references therein). On the down
side, the traditional composition had led to a truly remarkable
agreement between the sound speed predicted by solar interior models
and the helioseismical inversions. A revision of the oxygen abundance,
along with the cascading effects on nitrogen, carbon and neon, would
require a thorough reworking of that comparison and some recent work
(\citeNP{DP06}; \citeNP{BCE+07}) already argues for its
incompatibility. The far-reaching implications of the proposed change
(notice that it represents a factor of two in the actual number
densities) has prompted \citeN{APK06} to label this problem as {\it
the solar oxygen crisis}.

In their work, \citeN{APK06} derive a semiempirical 1D model of the
average thermal stratification and use it to obtain the oxygen
abundance from observed CO bands. This approach led them to
recommend a rather high value of \lgEO=8.85, consistent with the {\it
  traditional} determinations. While discussing the discrepancy with
the \citeN{AGS+04} results, they correctly point out that a 3D
theoretical simulation is not necessarily superior to a 1D
semiempirical model when used for the diagnostics of
observations. They both have advantages and drawbacks. 

The approach taken here tries to capture the best elements of both
worlds, by building a semiempirical 3D model of the solar
photosphere. The data set was acquired on 2006 Oct 27 with the
Spectro-Polarimeter for Infrared and Optical Regions (SPINOR,
\citeNP{SNEP+06}). It consists of a high-resolution (0.7'')
spectro-polarimetric scan of a quiet Sun region including two \FeI \
lines at 6302~\AA \ and the oxygen infrared triplet at
7774~\AA . The \FeI \ lines have been used to build the semiempirical
3D model, while the \OI \ triplet is employed for the abundance
determination. Since both sets of lines have been observed
simultaneously, there is full consistency between the data employed to
construct the model and to derive the abundances.

\begin{figure}
\plotone{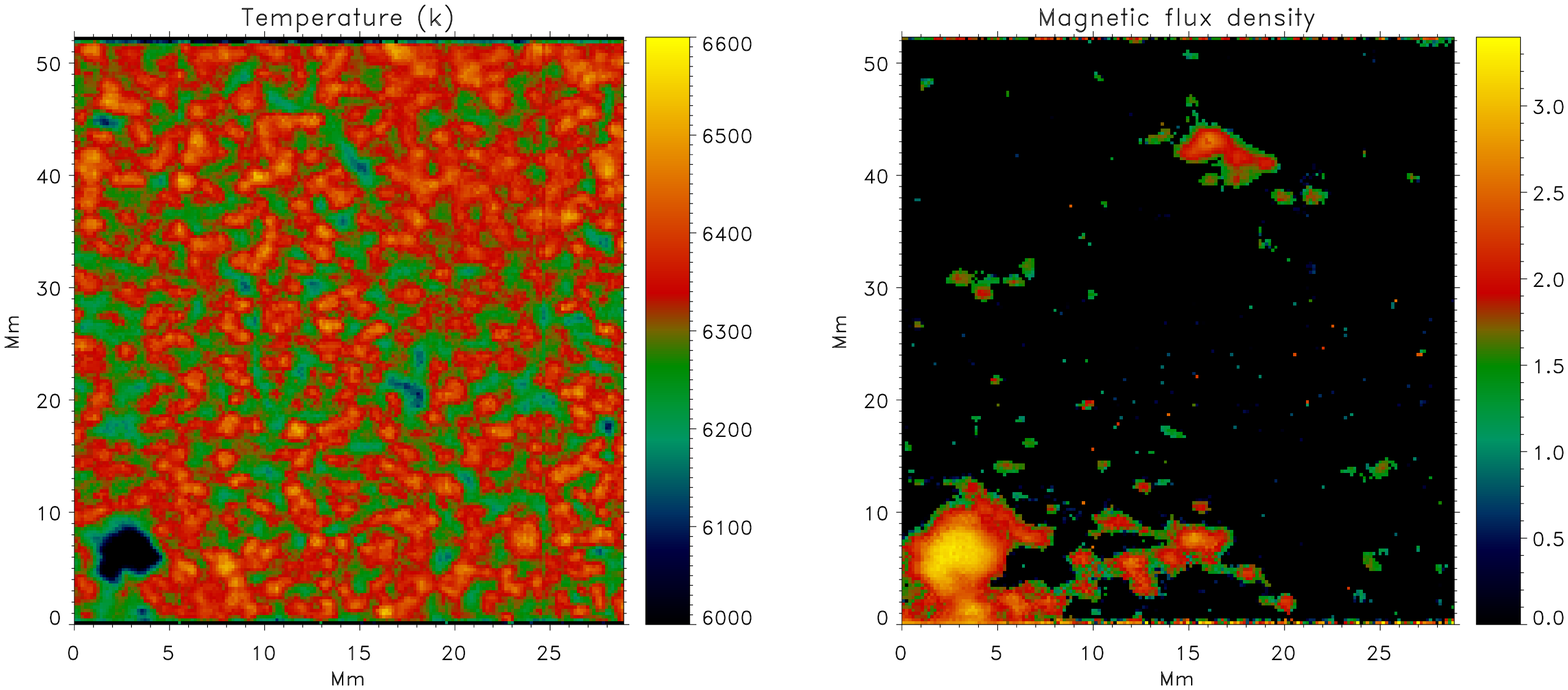}
\caption{Maps of the temperature and the magnetic flux density at the
  base of the photosphere in the 3D model, as derived from the SPINOR
  data. Left panel: Temperature. In those pixels that exhibit
  polarization signal, the average temperature inside and outside the
  magnetic concentration (weighted according to occupation fraction)
  is depicted.
  The cool feature on the lower left corner is a pore that was
  used to stabilize the adaptive optics, allowing us to achieve the
  spatial resolution shown in the image. Right panel: Decimal
  logarithm of the magnetic flux density in G.
  \label{fig:maps1}
}
\end{figure}

Our 3D model has been obtained from the inversion of the well-known
\FeI \ lines at 6301.5 (for which the oscillator strength has been
measured in the laboratory, \citeNP{BKK91}) and 6302.5 \AA \ together
with the continuum at 7774 \AA . For each pixel in the field of view,
we derived the vertical stratification of temperature, density,
line-of-sight velocity and longitudinal magnetic field (details on the
\FeI \ observations and the inversion will be provided elsewhere). We
impose the condition that the average observed quiet Sun continuum
intensity matches that of the Harvard Smithsonian Reference Atmosphere
(HSRA, see \citeNP{GNK+71}).  Although microturbulence is permitted as
a free parameter, the inversion code always retrieved very small
values ($\sim$100~m~s$^{-1}$). The model is publicly available upon
request from the authors. The spatial resolution is obviously limited
when compared to that of the Asplund et al. simulations, but it is
sufficient to resolve the granulation. Moreover, those pixels
exhibiting siginficant polarization signal have been treated with a
two-component scenario (one for the magnetic concentration and another
for the non-magnetic surroundings). Fig~\ref{fig:maps1} shows a
composite image of the temperature and the magnetic flux density at
the base of the photosphere.

The oxygen abundance was determined independently at each pixel of the
map, using the model atmosphere from the \FeI \ inversion, and
accounting for magnetic fields (i.e., considering the Zeeman splitting
and solving the Stokes vector transfer equation) where
appropriate. Synthetic profiles were computed at intervals of 0.1~dex,
and the $\chi^2$-vs-\lgEO \ curve was interpolated to find the minimum
with $\sim$0.01~dex accuracy (the $\chi^2$ curves obtained are very
smooth). For the LTE calculations we used the Stokes synthesis code
LILIA (\citeNP{SN01a}), whereas for the non-LTE case we employed the
code developed by \citeN{SNTBRC00a}. To alleviate the computational
burden we used a 1.5D approach (i.e., neglecting lateral radiative
transfer). This approximation is justified by \citeN{AGS+04}, who
obtained very similar results when comparing 3D and 1.5D syntheses
in their simulation. The model atom considered has 13 \OI \ levels
plus the continuum, with 18 bound-bound transitions and 13 bound-free
(6 of them treated with radiation temperatures). The atomic data are
the same as in \citeN{CJ93}.

Following the recommendation of \citeN{AGS+04}, we neglect excitation
and ionization due to inelastic collisions with neutral hydrogen. For
the collisional broadening of the lines we employed the formulation of
\citeN{G76} and tested the influence of the Van der Waals enhancement
parameter on the oxygen abundance. For this test we took the
spatially-unresolved disk center atlas of \citeN{NL84}, inverted the
\FeI \ pair of lines around 6302~\AA \ and used the resulting 1D model
to synthesize the \OI \ IR triplet. Varying the damping enhancement
parameter between 0 and 10 had very little influence on the oxygen
abundance ($\sim$0.01~dex). For larger values, the synthetic profiles
are too broad to fit the observations. While the enhancement parameter
strongly influences the equivalent widths of the lines (because it
broadens the wings leaving the cores more or less intact), its effect
is less pronounced in our analysis that considers the least-squares
fit to the overall profile.  In this test with unresolved observations
and a 1D model we obtained an oxygen abundance of \lgEO=8.66, which is
consistent with the previous 1D NLTE determinations of \citeN{AGS+04}
and \citeN{STBA05}.

\begin{figure}
\plotone{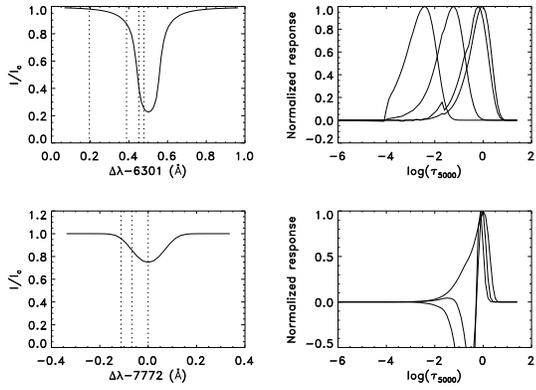}
\caption{Normalized response functions to temperature perturbations,
  computed in the HSRA quiet Sun model. Left panels: Line profiles of
  the \FeI \ 6301 \AA \, (top) and \OI \ 7772 \AA \, (bottom) lines. Right
  panels: Response of the intensity profile at the wavelengths marked
  with vertical dotted lines in the left panels, as a function of
  height where the perturbation occurs. The 7772 \AA \, line is the
  deepest (darkest core) of the multiplet and therefore spans the
  broadest height range of the \OI \ triplet. Nevertheless, its formation
  range is still well within that spanned by the \FeI \ lines.
  \label{fig:rfs}
}
\end{figure}

Since the \OI \ lines are shallower than the \FeI \ lines, one would
intuitively expect their formation height range to be narrower. In
other words, the height range that we need for the synthesis of the
\OI \ triplet is contained entirely in the range that we can reconstruct
from the \FeI \ lines. A formal verification of this is presented in
Fig~\ref{fig:rfs}, which shows the temperature response functions
(RFs, see \citeNP{LdILdI77}) of the lines at several wavelengths, from
the continuum to the line center. RFs measure how the line profile
reacts to atmospheric perturbations at various heights, and therefore
provide very valuable information on the sensitivity range of spectral
features. 

\begin{figure}
\plotone{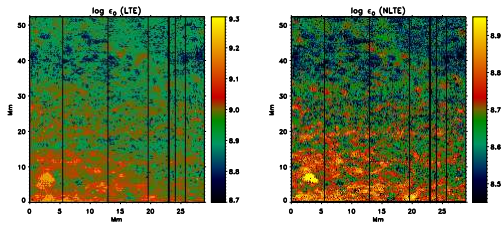}
\caption{Logarithmic oxygen abundance \lgEO \ (with the usual
  convention in Astrophysics that $\log \epsilon_{\rm H}$=12). Left
  panel: LTE calculation. Right panel: NLTE calculation. Note that the
  color scales are different, since the LTE abundances are
  systematically higher. The vertical lines are missing data due to
  errors in the detector system that was recording the oxygen data.
  \label{fig:maps2}
}
\end{figure}

The spatial distribution of \lgEO \ obtained with our 3D semiempirical
model is shown in Fig~\ref{fig:maps2}. In agreement with previous
works, we find that the LTE approximation largely overestimates \lgEO
\ by approximately 0.3~dex (this value is slightly dependent on
spatial location). The variation of the inferred \lgEO \ over the
field of view is due to fluctuations in the solar atmospheric
conditions. We can see that the abundance in the pore is considerably
higher, reaching values up to 0.3~dex larger than the average, and
decreasing gradually as we move away from it. In general, magnetic
concentrations exhibit higher abundances. This suggests that previous
abundance determinations might be slightly biased by solar (or, more
generally, stellar) activity, depending on the phase of the cycle in
which the observations were taken. The mean values that we obtained
over the entire field of view are \lgEO =8.94 (LTE) and \lgEO =8.64
(NLTE) dex, with standard deviations of 0.08~dex. In the NLTE case,
this is equivalent to a concentration of $\epsilon_O$/$\epsilon_H$ of
442 ($\pm$80) $\times$10$^{-6}$. If we restrict the analysis to pixels
with less than 100~G of flux density, we obtain \lgEO =8.93 (LTE) and
\lgEO =8.63 (NLTE).

Obviously, there is no physical reason to expect the actual abundance
to exhibit spatial variations in the solar photosphere. We must then
conclude that this is an artifact of the analysis, probably due to
imperfect modeling especially in the presence of magnetic fields
(notice that the granulation pattern is not visible in the abundance
images). We stress that the modeling employed here should be more
reliable than that in previous works (since it considers 3D geometry,
magnetic fields, NLTE and is based on actual observations). Yet, it
still falls short of providing a solid, accurate, abundance
determination. The uncertainties ($\sim$0.1~dex) are larger than
previously thought (\citeNP{PD07} have already pointed this out and
their error budget is consistent with our figure), and
this is also an important point to bear in mind.

\acknowledgments 
The authors are grateful to the NSO staff at Sunspot
NM (USA), and particularly the observers, for their support of the
SPINOR observations presented here. Thanks are also due to the
anonymous referee who helped improve a previous version of the
manuscript. HSN also acknowledges partial support from the Spanish
Ministerio de Educaci\'on y Ciencia through project
AYA2004-05792. Some of the processing power used for the calculations
in this paper has been generously donated by B. Lites, M. Wiltberger,
M. Rempel and J. Borrero.


\clearpage

\clearpage

\clearpage

\end{document}